\documentstyle[emulateapj,flushrt,epsf]{article}

\def \approxgt{\,\raise2pt \hbox{$>$}\kern-8pt\lower2.pt\hbox{$\sim$}\,}
\def \approxlt{\,\raise2pt \hbox{$<$}\kern-8pt\lower2.pt\hbox{$\sim$}\,}
\def \th{\thinspace}
\def \ngth{\negthinspace}
\def \ngth2{\negthinspace\negthinspace}
\def \ni{\noindent}
\def \Teff{{$T_{\rm {ef\!f}} $}}
\def \teff{{T_{\rm {ef\!f}} }}
\def\Lo{{$L_\odot $}}
\def\Mo{{$M_\odot $}}
\def \at{{\rm\char'100}}

\def\ah{{\scriptscriptstyle \frac 12 }}
\def\aq{{\scriptscriptstyle \frac 14 }}

\def \eg{{{\it e.g.},\ }}
\def \etal{{\it et al.\ }}

\def \cf{{\it cf.\ }}

\def \ie{{{\it i.e.},\ }}

\def \viz{{\it viz.\ }}

\def\Log{{\mathrm Log}}

\begin{document}

\submitted{ASTROPHYSICAL JOURNAL, in press}

\title{Mode Switching Time Scales in the Classical Variable Stars}
\author{J. Robert Buchler$^{1}$ \&
        Zolt\'an Koll\'ath$^{2}$}

\begin{abstract}

Near the edges of the instability strip the rate of stellar evolution is larger
than the growth-rate of the pulsation amplitude, and the same holds whenever
the star is engaged in pulsational mode switching.  Stellar evolution therefore
controls both the onset of pulsation at the edges of the instability strip and
of mode switching inside it.  Two types of switchings (bifurcations) occur. In
a {\sl soft} bifurcation the switching time scale is the inverse harmonic mean
of the pulsational modal growth-rate and of the stellar evolution rate.  In a
{\sl hard} bifurcation the switching times can be substantially longer than the
thermal time scale which is typically of the order of a hundred periods for
Cepheids and RR Lyrae.  We discuss some of the observational consequences, in
particular the paucity of low amplitude pulsators at the edges of the
instability strip.

  \end{abstract}


\date{\today}

\keywords{Stellar Evolution
-- Instabilities
-- Oscillations of stars - \th Cepheids - \th RR Lyrae 
}

 {\smallskip
        {\footnotesize
 \noindent $^1$Physics Department, University of Florida, Gainesville, FL, USA;
 buchler\at phys.ufl.edu \\
 \noindent $^2$Konkoly Observatory, Budapest, HUNGARY; kollath\at konkoly.hu
 }}

\section{Introduction}

\hyphenation{Stelling-werf}

The Cepheid and RR Lyrae instability strips (ISs) are observed to be rich in
behavior, as seen for example in the beautiful work of Udalski \etal (1999),
with stars undergoing single mode (SM) pulsations in the fundamental (F) mode,
in the first (O1) or second overtone (O2), or beat pulsations in two modes
(also called double-mode pulsations, DM).  Theoretical work has shown that for
the Galactic Cepheids we have a good understanding of the F and O1 Cepheids
(Moskalik, Buchler, Marom 1992, Bono, Marconi \& Stellingwerf 2000,
Feuchtinger, Buchler \& Koll\'ath, 2000) and of RR~Lyrae (Feuchtinger 1999).
Furthermore the problem of DM Cepheids and RR Lyrae has finally been solved
(Koll\'ath, Beaulieu, Buchler \& Yecko 1998, Feuchtinger 1998, Koll\'ath,
Buchler, Szab\'o \& Csubry, 2002, hereafter KBSC, Koll\'ath \& Buchler, 2001,
hereafter KB).  However some small problems remain, in RR Lyrae (Koll\'ath,
Buchler \& Feuchtinger 2000) and Cepheids, in particular those with low
metallicity (\eg Feuchtinger \etal 2002).

As the stars navigate through the IS their structure undergoes changes.
Accordingly, the allowed stable pulsational states change on an evolutionary
time scale, and the stars need to switch their pulsational behavior.  During
some stages more than one type of pulsational behavior is possible.  Some of
these pulsational states may be reachable, while other may not be (because a
finite amplitude kick would be required).  Stars can therefore be in a
different pulsational state at the same place on the blueward and the redward
tracks (hysteresis), as for example in the case of Cepheids and RR ~Lyrae.
Some classical Cepheids and RR~Lyrae have been observed to change their
pulsational status on human time scales.  For example, the pulsation amplitude
of Polaris has decreased dramatically over several decades (\eg Kamper \&
Fernie 1998).  Changes have also occurred in RR Lyrae variables (\eg Clement \&
Goranskij 1999).  The theoretical aspects of these mode switchings have been
addressed in the past (\eg Stellingwerf 1975, Bono, Castellani \& Stellingwerf
1995), but many of the questions that observations pose have remained
unanswered.  For example, observations indicate a paucity of low amplitude
s--Cepheids 
fundamental Cepheids  (\eg Fernie \etal 1995 for the Galaxy, 
Beaulieu, private comm.  for LMC and SMC), and 
no stars with low component amplitudes in the
DM region (Beaulieu, priv. comm., see figure 4 in Buchler 1998).  A similar
situation occurs for RR Lyrae stars (\eg Walker \& Nemec 1996, Walker 1998,
Udalski \etal 1997).  These deficiencies of small amplitude pulsators seem too
large to be solely due to observational bias.

Here we address the problem of mode switching in a novel fashion, by taking
temporal evolution explicitly into account.  Hydrodynamical modelling, to be
feasible, has typically decoupled pulsation from evolution (usually called a
quasi-static approximation).  The decoupling allowed the calculation of all the
possible full amplitude and steady pulsational states of a given model at any
point along the evolutionary track, \ie of a sequence of models.  However, this
quasi-static approximation can break down near the bifurcation points where the
star has to switch from one behavior to another (\eg Lebovitz 1990).  (A
bifurcation point is where there occurs a qualitative change in the modal
behavior).  The reason is that, right at a bifurcation point, the growth-rate
of a mode or limit cycle can vanish and remain small in the
vicinity.  This causes a temporal delay for the pulsation to achieve the
full amplitude that the hydrodynamical modelling would suggest -- {\it the
pulsational behavior is then controlled by evolution.}  In the following we
make use of the amplitude equation formalism (Buchler \& Goupil 1984, Buchler
1993) which is perfectly tailored to the study of what is happening in the
vicinity of the modal bifurcation points, and we show that there are
observational consequences.

\vskip 20pt

\section{Analytical Considerations}

We will be led to distinguish two different types of mode switchings that we
label soft and hard.  A {\it soft} bifurcation occurs when the nascent mode
starts off with infinitesimal amplitude and the growth-rate is also
infinitesimal.  At a {\it hard} bifurcation the stable fixed point disappears
(the current pulsational behavior is no longer allowed) and the star is left
looking for another stable fixed point (\ie another pulsational behavior).  The
rate of change of the amplitude is generally {\it not} infinitesimally small in
this case.
In \S3 we will see examples of the two types of switchings.

\subsection{Soft Bifurcations}

\subsubsection{Hopf Bifurcation}

We first consider the simplest soft modal switching which occurs when the star
enters or exits the IS.  Without lack of generality let us
consider the neighborhood of the blue edge where the linear growth-rate
$\kappa$ of a mode, the fundamental or the first or second overtone depending
on the luminosity, changes from negative (vibrationally stable) to positive
(unstable).  Generically, in the vicinity of the blue edge $\kappa(t)$ varies
proportionately to time, $\kappa = \chi t$, where the origin of time is taken
to be the passage through the blue edge.  The quantity $\chi=d\kappa/dt$ is
therefore the stellar evolutionary rate of change of the linear growth-rate.
This type of bifurcation is well known to dynamicists 
and goes under the name of supercritical Hopf bifurcation 
(\eg Berg\'e \etal 1984).  The behavior of the
amplitude is governed by an {\it amplitude equation} of the form (\eg Buchler
\& Goupil 1984, Coullet \& Spiegel 1984) 
\begin{equation} 
{d\over dt}A = \kappa(t) \th A - q(t) A^3 \quad\quad {\rm with}\quad\quad 
\kappa(t) \th =\chi \th t 
\label{eq_hopf} 
\end{equation} 
\ni where $\kappa(t)$ is the instantaneous linear growth-rate of the mode and
$q(t)$ a nonlinear saturation term which varies little along an evolutionary
track.

If the star were perturbed away from its quasi-static stable
limit cycle amplitude
\begin{equation}
A_{qs}(t) = \sqrt{\kappa(t)/q(t)}
\label{eq_lc}
\end{equation}
(as obtained from setting $dA/dt=0$ in Eq.~\ref{eq_hopf}), it
would return to it on a $t_{th}$$\th\sim\th$
$\kappa^{-1}$ time scale, \ie over a few hundred
pulsations, because the relative growth-rates are typically of the
order of a percent inside the IS.
Away from the bifurcation, the thermal time scale $t_{th}$ 
is much shorter than the stellar evolution time scale
$t_{ev}$$\th\sim\th$$(d\ln\teff/dt)^{-1}$.  

Right at the bifurcation, however, $\kappa$ vanishes by definition, and it
remains small in some neighborhood, so that $t_{th} \gg t_{ev}$ for a while,
and the amplitude does not grow.  In fact the star must wait for evolution to
increase the value of $\kappa$.  Only then can the amplitude catch up with its
quasi-static value $A_{qs}(t)$.  It should be clear from this discussion that
{\it the evolutionary rate of change of the growth-rate $\chi$ will play a
crucial role}.

Let us call $\hat t$ the time when the solution catches up, and $\hat A$ the
amplitude at that instant.  We are interested in determining the dependence of
$\hat t$ and $\hat A=A_{qs}(\hat t)$ on $\chi$, $q$ and $A_{\rm o}$=$A(t_{\rm
o})$.  The latter is the pulsation amplitude at $t_{\rm o}$ that arises because
of noise in the star.

The coefficient $q$ is time-dependent because of changes in the structure of
the star as it evolves, but its time-dependence is small.  In a later example
of an actual RR Lyrae model we include its time-dependence.  Here, we
approximate
$q$ as a constant.  With the substitution $E=A^2$ we can then readily solve
Eq.~(\ref{eq_hopf}) in the form
\begin{eqnarray}
E(t)^{-1} & \simeq & e^{-\chi t^2}/E_{\rm o}
  + 2q \int_{t_{\rm o}}^t e^{-\chi (t^2-t'^2)} dt'  \nonumber \\
   &\simeq & e^{-\chi t^2}/E_{\rm o} + q/( \chi t)
 \label{eq_et}
\end{eqnarray}
\noindent where the last expression is the first term of an
asymptotic expansion for the integral for large $\chi t^2$. 

The point at which the pulsation amplitude catches up with
the quasi-static value can be obtained to a good approximation 
by requiring that 
the first term on the rhs of Eq.~(\ref{eq_et}), which completely dominates at
first, becomes a fraction $\xi$ of the second term ($\xi=0.1$ turns out to be a
good value).
With the help of Eq.~(\ref{eq_et}) this yields a simple transcendental 
equation
\begin{eqnarray}
\quad 2 u - \ln u - \ln f = 0\th, \quad {\rm where} \quad u = \chi t^2\th ,
\end{eqnarray}
\begin{equation}
{\rm and} \quad f = {\chi \over ( \xi q A_{\rm o}^2 )^2} .
\end{equation}
A good approximation to the solution, when f is large, is obtained with the
fitting expression
\begin{eqnarray}
\quad \hat u = \chi \hat t^{\th 2} \sim 1.2 \th\th \Log\th f + 0.7
\th .
\label{eq_approx}
\end{eqnarray}
Finally we have $\hat t  =  (\hat u/\chi)^\ah$.
The quantity q can be re-expressed in terms of the maximum pulsation amplitude
in the IS, since $q A_{max}^2 = \kappa_{max}$.  Thus
${\hat A/ A_{max}}  =  (\chi \hat t / \kappa_{max})^\ah$.

  \vspace{0.5cm}

  \centerline{\vbox{\epsfxsize=8.7cm\epsfbox{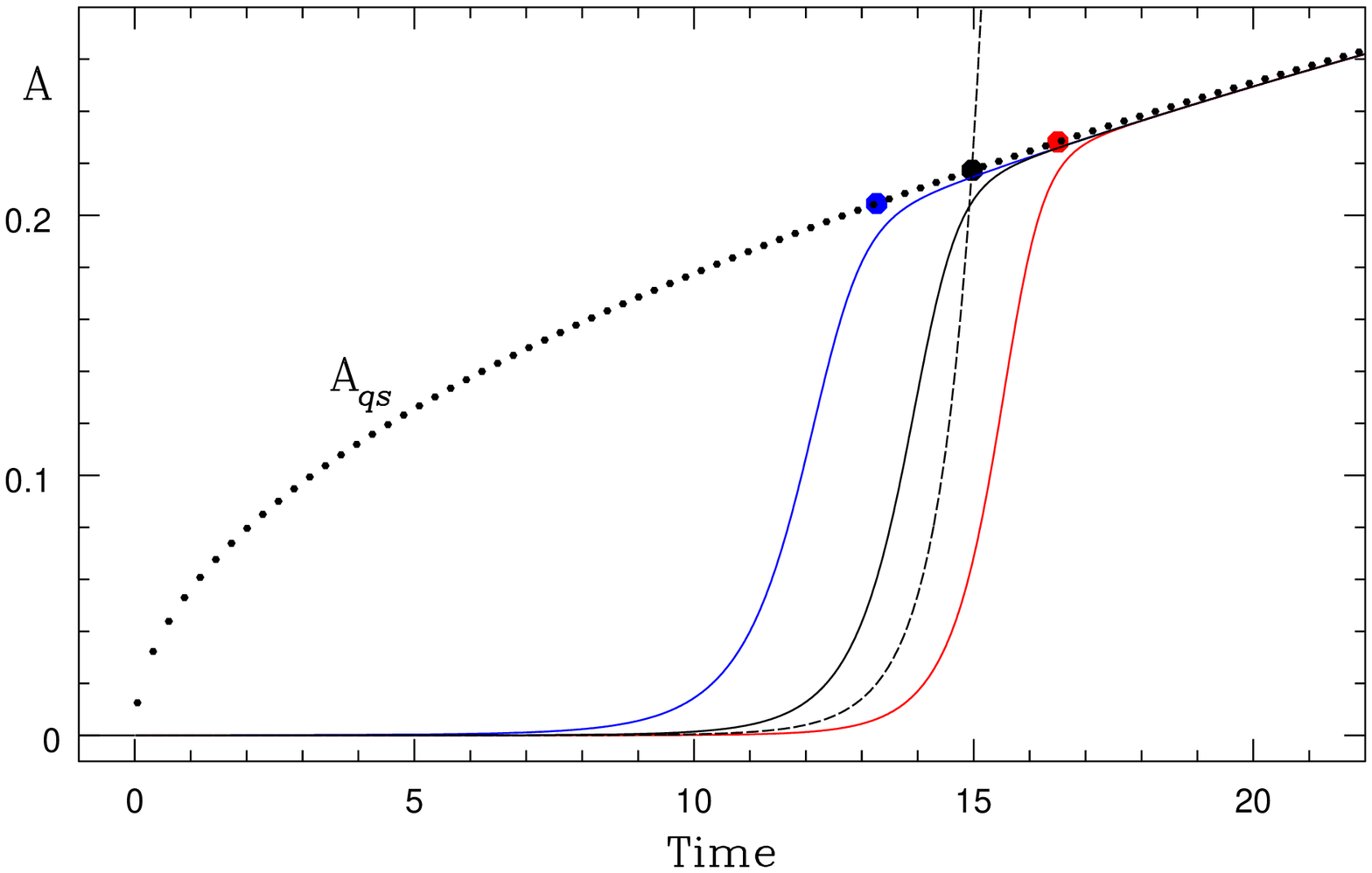}}}

  \noindent{\small Fig.~1: Temporal behavior of the pulsation amplitude after
the star enters the IS (solid line), for the case of $\chi = 10^{-16} s^{-2}$.
Curves are for three values of initial perturbation (noise level) ${\rm Log}\th
A_{\rm o} = -14, -16, -18$.  The dotted line represents the quasi-static
amplitude $A_{qs}(t)$ that would be achieved if the pulsation could react
infinitely fast to evolutionary changes in $\kappa$.  The amplitude scale is
set by the choice $q=10^{-6}$.  The thick dots correspond to the $\hat A$
obtained with the approximation of Eq.~6.  The dashed line represents 
$A_{\rm o} \exp(\ah\chi t^2)/\xi $ for the middle value of $A_{\rm o}$.}

  \vspace{0.2cm}


The typical behavior of $A(t)$ as computed from Eq.~(\ref{eq_hopf}) is plotted
as solid lines in Fig.\th 1 for ${\rm Log}\th \chi = -16$ and for three values
of $A_{\rm o}$ (from left to right: ${\rm Log}\th E_{\rm o}=$ --14, --16,
--18).  The quantity $q$ which scales the amplitude, has been arbitrarily set
to $q=10^{-6}$.  Shown as a dotted line is the quasi-static limit cycle amplitude
$A_{qs}(t) = \sqrt{\chi t /q}$, namely the value that the amplitude would
achieve if the system could react instantly.

We have no knowledge of the intensity of the fluctuation $A_{\rm o}$ in a star
which represents the projection of the turbulent noise on the eigenvector of
the mode.  As suggested by the approximation formula, the larger $A_{\rm o}$
is, the faster the amplitude grows (Fig.~1).  But one notes that the effect is
only logarithmic and therefore a precise knowledge of $A_{\rm o}$ is of little
importance.  {\it It is the rate of change of the growth-rate, $\chi$, that
sets the switching time scale}, \ie the scale of the abscissa in Fig.~1.

The three fat dots denote the $\hat t$ and $\hat A$ for the three values of
$A_{\rm o}$ as given by the fitting formula.  It can be seen
that they provide an excellent agreement with the numerical integrations of
Eq.~(\ref{eq_hopf}).  The dashed line represents $A_0 \exp(\ah\chi t^2)/\xi $
for the middle value of $A_{\rm o}$.

\subsubsection{SM to DM Bifurcation}

When two (nonresonant) excited modes are involved Eq.~~\ref{eq_hopf} needs to
be replaced by the two coupled amplitude equations for modes 1 and 2
(\eg Buchler \& Goupil 1984)
\begin{eqnarray}
\!\!\dot A_{\rm 1} \!\!\!&= \Big(\kappa_1(t) + q_{11}(t) A_{\rm 1}^2
              + q_{12}(t) A_{\rm 2}^2 + s_{1}(t) A_{\rm 1}^2 A_{\rm 2}^2
                         \Big) \th A_{\rm 1} \nonumber &(7a)\\
\nonumber \\
\!\!\dot A_{\rm 2}\!\!\! &= \Big(\kappa_2(t) + q_{21}(t) A_{\rm 1}^2
              + q_{22}(t) A_{\rm 2}^2 + s_{2}(t) A_{\rm 1}^2 A_{\rm 2}^2
                         \Big) \th A_{\rm 2}  \nonumber &(7b)
 \label{eq_aes}
\end{eqnarray}
We have kept some of the quintic terms in these equations which are necessary
in order to describe the observed modal behavior (KB, KBSC). \th (Earlier
studies by Buchler \& Kov\'acs (1987) had hoped to get away with retaining only
the cubic ones).

When the transition from SM to DM inside the IS is such that the nascent mode
starts with infinitesimal amplitude, as we assume in this section on soft
bifurcations, then the behavior of this amplitude is very similar to the Hopf
bifurcation that we have just described.  The reason is easy to see: in the
vicinity of the bifurcation the amplitude of the finite amplitude mode (say
$A_{\rm 1}$) changes very little, and we can expand around this point which
leads to an equation for the amplitude say $A_{\rm 2}$ of the nascent mode that
is identical to Eq.~\ref{eq_hopf}.

  \vspace{0.2cm}

To summarize then, stellar evolution causes a delay of order $\chi^{-\ah}$
during which the pulsation amplitude stays negligibly small, but after which it
very rapidly achieves its quasi-static value.  The time spent at the small
intermediate amplitudes is therefore very small and the stars are very unlikely
to be observed with such small amplitudes.  The amplitude at catch-up $\hat
A$$\th\sim\th$$\chi^\aq$ thus also depends on $\chi$, and when the stellar
evolution is slow it will be large.  Finally, we note that the time scale and
the catch-up amplitude are both insensitive to the poorly known noise level,
typified by $A_{\rm o}$.


\subsection{Hard Bifurcations}

A hard bifurcation occurs when, as a result of stellar evolution, a fixed point
merges with another unstable fixed point, in a mutual annihilation.  One can
readily visualize this in an amplitude--amplitude plane (see \eg KB, KBSC),
where the fixed points, stable and unstable, lie at the intersection of two
loci (curves).  As \Teff\ changes with evolution, the two intersection points
coalesce and then disappear as the curves move apart.  At that point the star
has no choice but to look for another stable pulsational state to mew into.
The new type of pulsation depends on the overall topography at the time, and is
determined by which fixed point attracts the stellar model.  One might think
that the time to attain the new pulsational state is the thermal time scale
$t_{th}=\kappa^{-1}$, but it can be and it is generally much longer as we shall
show in a realistic example.  We note that for a hard bifurcation the
stellar evolution time scale plays no important role because the amplitude
growth-rates do not vanish and thus the time scale for amplitude change is
always finite (KBSC).

\section{Cepheids and RR Lyrae}

\subsection{The Blue Edge}

Let us consider first a redward evolving 6 \Mo\ Cepheid that enters the first
overtone IS.  We use Eq.~(\ref{eq_approx}) to estimate the minimum observable
amplitude $\hat A$.  Stellar evolution calculations indicate that a 6\Mo\
Cepheid variable evolves at a rate of $d
\teff/dt$$\th\sim\th$$2\times$$10^{-10}$$K\th s^{-1}$ when entering the
overtone blue edge (\eg Alibert \etal 1999).  Linear models give a rate of
change $d\kappa/d\teff$$\th\sim\th$ $2\times$$10^{-10}$$(Ks)^{-1}$, so that we
obtain $\chi$$\th\sim\th$$4\times$$10^{-20}$$s^{-2}$.  The maximum growth-rate
$\kappa_{max}$ in the IS is $4\times$$10^{-8}$$s^{-1}$.  If we assume intrinsic
fluctuations of the order of $A_0/A_{max}$$\th\sim\th$$10^{-10}$, where 
$A_{max}$ is the maximum amplitude in the instability strip, we find that
the $\hat u$ in Eq. 6 takes on values around 35 -- 40.  Thus we estimate that
$\hat A/A_{max}$\th$\sim\th$0.1 at the blue edge, and that the pulsation
amplitude stays negligibly small for $\hat t$$\th\sim\th$1000\th yr, and that
it then increases very rapidly to a value 10 \% of the maximum amplitude in the
IS.  Because of this rapid increase of the amplitude very few overtone Cepheids
with amplitudes less than $\hat A\th\sim\th A_{max}$ should be observed near
the blue edge.

Turning now to a 8\Mo\ Cepheid which evolves at \break $d \teff
/dt$$\th\sim\th$$2\times$$10^{-9}$$Ks^{-1}$ \th(Alibert \etal 1999) when
entering the fundamental blue edge, with a
$d\kappa/d\teff$$\th\sim\th$$4\times$$10^{-11}$$(Ks)^{-1}$ we obtain
$\chi$$\th\sim\th$8$\times$$10^{-20}$$s^{-2}$.  The maximum $\kappa_{max}$ in
the IS is $\sim$\th 7$\times$$10^{-9}$$s^{-1}$.  Thus we estimate that again
$\hat A/A_{max}$$\th\sim\th$0.1 near the blue edge, and that the pulsation
amplitude stays negligibly small for $\hat t$$\th\sim\th$800\th yr.

Cepheids evolve both blueward and redward.  Only that half of the Cepheids near
the blue edge will therefore be affected, namely only those that move redward
and enter the IS.  For the same reasons, there should also be a (50\%)
deficiency of Cepheids at the red edge.  Furthermore at a given period there
will be a medley of Cepheids spread over over the whole IS.  Consequently there
should be an overall deficiency of low amplitude Cepheids at all periods.
  
\vskip 5pt

For RR~Lyrae stars we deduce from Dorman (1992) that $d \teff/dt$$\th\sim\th$
$1.2\times$$10^{-12}$$s^{-1}$ when entering the overtone blue edge.  From our
linear models we obtain
$d\kappa/d\teff$$\th\sim\th$$1.5\times$$10^{-9}$$(Ks)^{-1}$ which leads to
$\chi$$\th\sim\th$$1.8\times$$10^{-21}$$s^{-2}$.  In the IS $\kappa_{max}$ is
$71\times 10^{-9}$$s^{-1}$.  We thus estimate that $\hat
A/A_{max}$$\th\sim\th$0.01 at the blue edge, and that the pulsation amplitude
stays negligibly small for $\hat t$$\th\sim\th$ 5000\th yr.

\subsection{Switching to and From Double-Mode Pulsation}

The coefficients of the amplitude equations (\ref{eq_aes}) can be extracted
from hydrodynamical simulations, as we will illustrate  for
a sequence of RR Lyrae models.   First we compute the stable pulsational states
at selected values in the \Teff\ range of interest.  From these
we extract for each model (characterized by \Teff) the coefficients
$\kappa_0(\teff)$, $q_{00}(\teff)$, $\ldots $.  The values of $d\ln T_{\rm
ef\!f} /dt$ along the evolutionary track through the IS determine the
time-dependence of $\kappa_0(t)$, $q_{00}(t)$, $\ldots$\ .

\vskip 0.4cm


  \centerline{\vbox{\epsfxsize=8.9cm\epsfbox{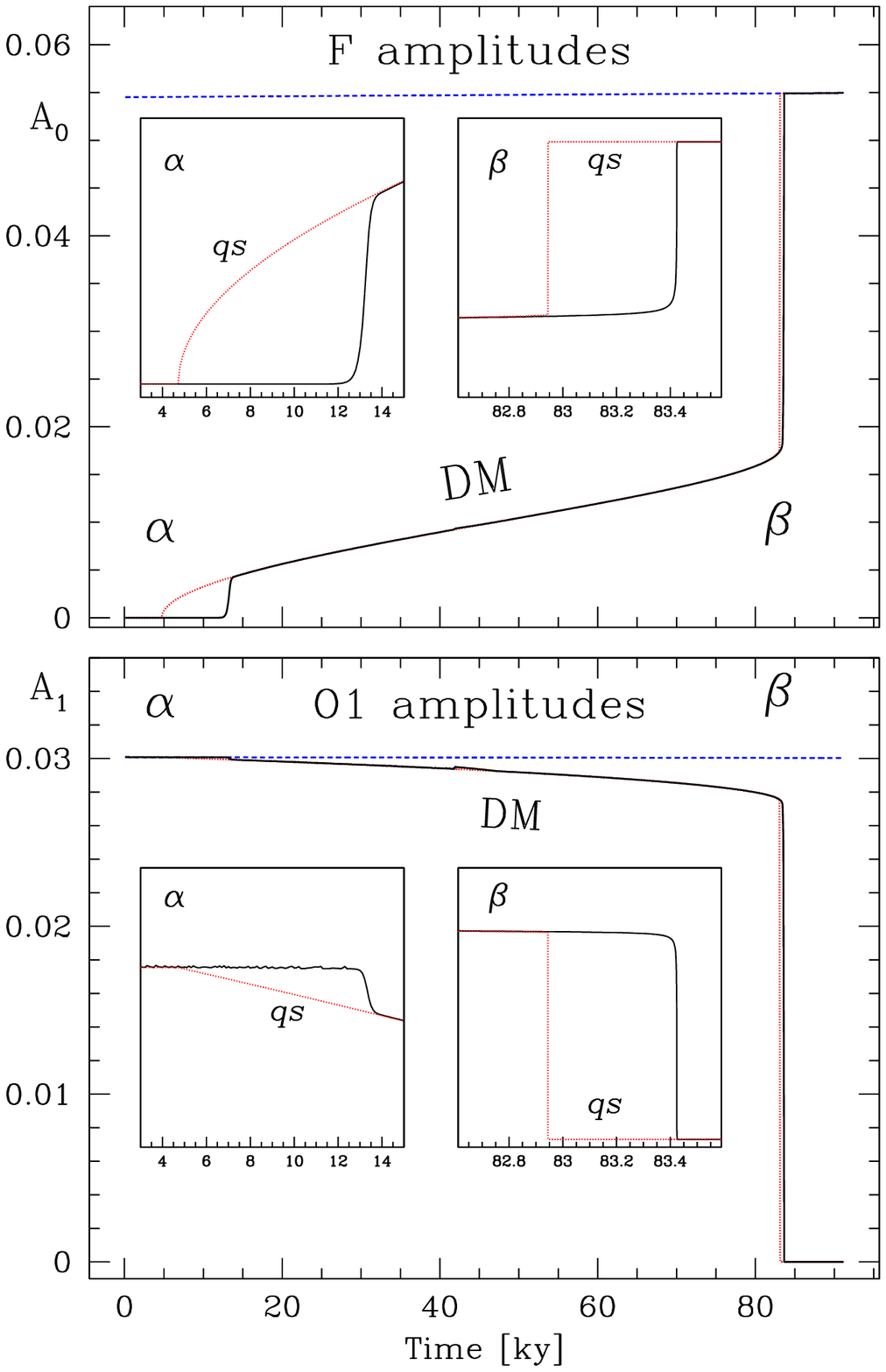}}}

  \vspace{0.2cm}

  \noindent{\small Fig.~2: 
Redward evolution of the RR Lyrae model through the HR diagram.
{\sl Top:} fundamental (F) SM amplitude or component amplitude in the DM;
{\sl bottom:} O1 SM amplitude or component amplitude.
At ($\alpha$): switching from SM
O1 to DM pulsation and at ($\beta$): switching from DM to the F
limit cycle.}

  \vspace{0.4cm}


We illustrate the method for an RR Lyrae star as it evolves through the
switching from SM overtone to DM to SM fundamental pulsation.  Figure~2 shows
the results of the integration of Eqs.~(\ref{eq_aes}), \ie the pulsation
amplitudes of the F and O1 modes.  For convenience we have chosen a track with
fixed luminosity L= 50\Lo\ and mass M= 0.77\Mo, Z= 0.0001, rather than an
inclined evolutionary track (Dorman 1992).  The star is assumed to be evolving
toward colder \Teff.  The origin of time is arbitrarily taken to be a few
thousand years before the point (marked ${\bf \alpha}$) where the O1 limit
cycle becomes unstable to a DM pulsation (point $\beta$).


Fig.~2 is best understood in conjunction with paper KBSC.    (Their
Fig.~3 shows the location of the SM and DM fixed points and their stability in
an amplitude--amplitude plot for selected
\Teff s, while their Fig.~2 shows the variation of the quasi-static
amplitudes along the sequence).
The solid lines here denote the pulsation amplitudes, or the component
amplitudes in the case of DM pulsations, as a function of time, top graph for
the F amplitude and bottom for the O1 amplitude.  The dotted lines denote the
quasi-static amplitudes, \ie the instantaneous fixed points of
Eqs.~(\ref{eq_aes}).

\vskip 0.4cm


  \centerline{\vbox{\epsfxsize=8.cm\epsfbox{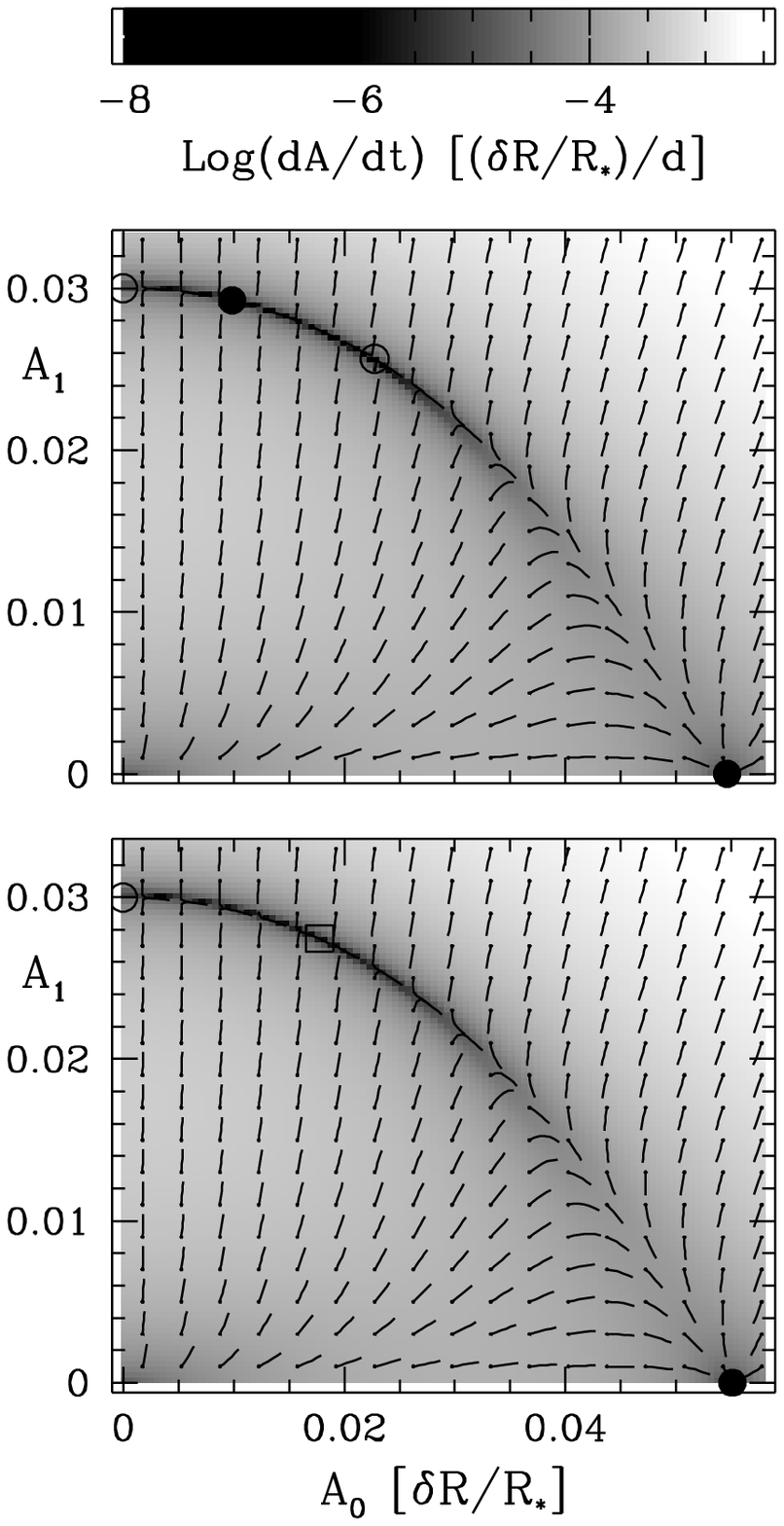}}}

\vskip 0.1cm

\noindent{\small Fig. 3: F amplitude -- O1 amplitude phase space for two RR
Lyrae models with a 10K difference in effective temperature, where the upper
one has the higher temperature.  The short lines represent the normalized flow
field (with the dots representing the bottom of the vectors).  The large
solid/open circles in the top graph denote the location of the stable/unstable
fixed points.  The square in the bottom denotes the location of the just
vanished pair of DMs.  The speed of the flow is represented by a gray scale.
The amplitudes refer to $(\delta R/R)_{\rm sur\!f}$.  }

  \vspace{.3cm}


\subsubsection{Soft Bifurcation}

The first inset in Fig.~2 is a blow-up of the onset of DM behavior near point
(${\bf \alpha}$).  As expected, the behavior of the nascent F amplitude is very
similar to the solution of Eq.~(\ref{eq_hopf}) that was shown in Fig.~1.  The
major difference is that here $\chi$ becomes the rate of change of the linear
stability coefficient of an F mode perturbation of the O1 limit cycle (the
Floquet stability exponent) rather than that of the linear growth-rate of the
mode.

One sees that, for DM onset as well, stellar evolution inhibits the growth of
the pulsation amplitude for some 8000\th yr, \ie much beyond the thermal time
scale which is of the order of a hundred days.  
DM RR~Lyrae (RRd) with
small fundamental mode component pulsation amplitudes should be rare.

\subsubsection{Hard Bifurcation}

An example of a hard bifurcation occurs in the second bifurcation for our RR
Lyrae model, marked ${\bf \beta}$ in Fig.~2.  The stable DM is annihilated by a
second unstable DM and they both disappear (\cf Figs.~2 and 3 of KBSC).  In
this case the star gets attracted by the F fixed point, and thus the pulsation
switches to a stable F limit cycle.

Figure~3 depicts the nature of the two RR Lyrae stellar models of Fig.~1, taken
on either side of point $\beta$, just before (upper panel) and after (lower
panel) the DM fixed point vanishes.  In this amplitude-amplitude diagram the
large solid circles represent the stable fixed points, of which now only the F
fixed point is left in the lower figure with amplitude $A_{\rm 0}$= 0.055.  The
large open circle refers to the unstable O1 fixed point at $A{\rm _1} =
0.030$. The large square denotes the location of the just vanished pair of DMs.

The 'velocity' flow field in the $A_{\rm 0}$--$A_1$ plane is given by ${\bf u}
= (\dot A_{\rm 0},\dot A_1)$, and the speed by $u=(\dot A_{\rm 0}^2 + \dot
A_1^2)^\ah$.  The 'arrows' in Fig.~3 represent the normalized 'velocity' flow
field , \ie the vectors ${\bf u}/u$.  The dots denote the bases of the vectors.
The gray scale indicates the Log of the speed (length of the velocity vectors).
Note that the typical 'thermal' time scale is given by the shading near the
origin.

It is very apparent that all the trajectories get attracted very rapidly to an
arc going from the $A_1$ axis to the the $A_{\rm 0}$ axis.  This arc is an
integral curve of the steady state amplitude equations (known as a heteroclinic
connection because it connects two different fixed points, \viz the O and F
fixed points.)  Along this arc the motion is considerably slower than thermal.
The transition occurs on a time scale of $\th\sim\th$100\th yr, very long
compared to the thermal time scale of $\kappa^{-1}$$\th\sim\th$100\th d.

At the moment the DM disappears, the star finds itself at the point marked with
a large square (bottom graph), and it gets slowly attracted to the only stable
fixed point, namely F.  The pulsation slowly changes from DM to fundamental SM.

In the case of a hard bifurcation, the switching time, even though of the 'same
order of magnitude' as the thermal time scale, thus can exceed it by factors of
ten.

\subsubsection{Speed Along Heteroclinic Connection}

Finally we examine why the speed is slow along the heteroclinic connections
that link the SM and the DM fixed points.  In Fig.~3 one visualizes the
location of these heteroclinic connections from the vector field.  (They are
shown explicitly in Fig.~1 of KBSC.)  Consider now the two loci that are
defined by setting the first and second of eqs.~(\ref{eq_aes}) equal to zero,
respectively.  On these loci either the horizontal ($\dot A_1$) or the vertical
($\dot A_2$) speed vanishes by definition.  These loci each emanate from one of
the two SM fixed points, and both of them intersect the heteroclinic connection
at the two DM points.  The heteroclinic connection lies therefore very close to
the two loci, especially near and in between the DM fixed points, and the speed
$u = (\dot A_1)^2 + (\dot A_2)^2)^\ah $ must therefore also remain small.

One also notes that, when both modes are linearly unstable, which is the
situation we are interested in, the heteroclinic connection is always
attractive sideways.  This remains true {\it a fortiori} when the loci
intersect and give rise to a DM fixed point (in which case the speed vanishes
at the DM).

Note also that the linear stability root of the F fixed point along the $A_{\rm
0}$ axis is given by $-2\kappa_0$ and, since $\kappa_0$ is typically large the
speed along $A_{\rm 0}$ is high.  {\it Mutatis} {\it mutandis} one shows that
the speed along the $A_{\rm 1}$ axis is also high.  Although this only concerns
the motion along the coordinate axis, Fig.~3 shows that it remains true in
between.


\section{Conclusion}

We have shown that there are two types of bifurcations with different
characteristics.  The first is a soft bifurcation where the incipient mode
starts with infinitesimal amplitude whose growth is determined by the
evolutionary rate of change $\chi = d\kappa/dt$ of the appropriate linear
growth-rate $\kappa$.  This time scale $\chi^{-\ah}$ can be very large compared
to the thermal time scale $\kappa^{-1}$ depending on the stellar evolution time
scale.  The second is a hard bifurcation which occurs when the rug is pulled
from under the pulsation, as it were, \ie when the pulsation must change,
because, due to an evolutionary change in the star's structure, it is no longer
allowed or possible.  Here the time scale remains thermal, but generally with a
large multiplication factor.  This delay and then abrupt onset of pulsation has
as a consequence that stellar pulsators with low amplitudes should be rarely
observed, in agreement with observations.

Another interest of this paper is that it shows the possibility of
obtaining valuable information about stellar structure from the observations of
a star that is in the process of starting to pulsate or of switching its
pulsational state.  In a subsequent paper we will show how the combined use of
numerical hydrodynamics, time-series analysis, the amplitude equation formalism
and the knowledge of the stellar evolution time scale (d\th T$_{\rm ef\!f}$
/dt) allow one to compute consistently the evolution of a pulsating star
throughout the whole IS.

\vspace{2mm}

This work has been supported by NSF (grant AST 9819608) and by OTKA (T-038440).
We also wish to thank an anonymous referee for his suggestions.


{}

\begin{thebibliography}{}


\bibitem[]{}
Alibert, Y., Baraffe, I., Hauschildt, P., Allard,
F. 1999 AA 344, 551

\bibitem[]{}
Berg\'e, P., Pomeau, Y., Vidal, C. 1986, {\it Order Within Chaos}
(N.Y. : Wiley)

\bibitem[]{} Bono, G., Castellani, V., Stellingwerf, R.F. 1995, ApJ 445, 145

\bibitem[]{} Bono, G., Marconi, M., Stellingwerf, R.F. 2000, AA 360, 245

\bibitem[]{} Buchler, J.R. 1993, in {\it Nonlinear Phenomena in Stellar
Variability}, Eds. M. Takeuti \& J.R.  Buchler (Kluwer: Dordrecht), repr. from
ApSS 210, 1

\bibitem[]{} Buchler, J.R. 1998, in ASP Conf. Ser. 135, 220

\bibitem[]{}
Buchler, J.R., Goupil, M. J. 1984, ApJ, 279, 394 


\bibitem[]{}
Buchler, J.R., Kov\'acs, G. 1987, ApJ 318, 232 

\bibitem[]{}
Clement, C.M., Goranskij, V. P. 1999, ApJ 513, 767

\bibitem[]{}
Coullet, P., Spiegel, E. A. 1984, SIAM J. Appl. Math., 43, 776


\bibitem[]{}
Dorman, B. 1992, ApJ Suppl 81, 221

\bibitem[]{}
Fernie, D.J., Beattie, B., Evans, N.R. \& Seager, S. 1996, IBVSNo 4148
(ddo.astro.utoronto.ca/cepheids.html)

\bibitem[]{}
Feuchtinger, M.U. 1998, AA 337, L29 

\bibitem[]{}
Feuchtinger, M.U. 1999, AA 351, 103 

\bibitem[]{}
Feuchtinger, M., Buchler, J.R., Koll\'ath, Z. 2000, ApJ 544, 1056 

\bibitem[]{} 
Feuchtinger, M., Buchler, J.R., Koll\'ath, Z., Beaulieu J.P. 2002, 
in preparation

\bibitem[]{}
Kamper, K.W., Fernie, J.D. 1998, AJ 116, 936 

\bibitem[]{}
Koll\'ath, Z., Beaulieu, J. P., Buchler, J.R. \& Yecko, P. 
1998, ApJLett 502, L55
 
\bibitem[]{}
Koll\'ath, Z., Buchler, J.R. 2001, {\sl Double-Mode Stellar
Pulsations}, in {\sl Nonlinear Studies of Stellar Pulsation},
Eds. M. Takeuti \& D.D. Sasselov, ApSS Lib Ser, Kluwer Vol. 257, p. 29,
(astro-ph/0003386), [KB].

\bibitem[]{}
Koll\'ath, Z., Buchler, J.R., Feuchtinger, M. 2000, ApJ  540, 468 

\bibitem[]{}
Koll\'ath, Z., Buchler, J.R., Szab\'o, R.,  Csubry, Z. 2002, AA
(in press), \th(astro-ph/0110076), [KBSC].

\bibitem[]{}
Lebovitz, N. 1990, in {\it Nonlinear Astrophysical Fluid Dynamics}, Eds.
    J.R.  Buchler \& S.T. Gottesman, Ann. NY Acad. Sci. 617, p. 73

\bibitem[]{}
Moskalik, P., Buchler, J.R., Marom, M. 1992 ApJ
385, 685

\bibitem[]{}
 Stellingwerf, R.F. 1975, ApJ 195, 441

\bibitem[]{}
 Udalski et al. 1997, Acta Ast. 47, 1.

\bibitem[]{}
 Udalski et al. 1999, Acta Ast. 49, 1.

\bibitem[]{}
Walker, A. 1998, AJ 116, 1005.

\bibitem[]{}
Walker, A. \& Nemec, J.M. 1996, AJ 112, 2026

\end{thebibliography}
\end{document}